# STOP THAT SUBVERSIVE SPREADSHEET !

David Chadwick

*School of Computing and Mathematical Sciences,*

*University of Greenwich, London SE10 9LS, UK;     cd02@gre.ac.uk*

Abstract:   This paper documents the formation of the European Spreadsheet Risks Interest Group (EuSpRIG www.eusprig.org) and outlines some of the research undertaken and reported upon by interested parties in EuSpRIG publications.

## 1.  EUSPRIG

The European Spreadsheet Risks Interest Group (EuSpRIG) was founded in March 1999 as a collaboration between spreadsheet researchers at the University of Greenwich, the University of Wales Institute Cardiff and HM Customs & Excise. Its mission was to bring together academics, professional bodies and industry practitioners throughout Europe to address the ever-increasing problem of spreadsheet integrity. EuSpRIG now has approximately 200 members. These are mainly individuals but there are also corporate members such as KPMG, PricewaterhouseCoopers, Superdrug, banks such as Lloyds TSB and the Netherlands Bank, and universities such as those of Hawaii, Amsterdam, Singapore, Klagenfurt, Calgary and Dartmouth College (USA). EuSpRIG also has close contacts with producers of audit software such as Operis (OAK), Southern Cross Software (Spreadsheet Detective), and HM Customs & Excise (SPACE).

EuSpRIG ([www.eusprig.org](www.eusprig.org)) has held four conferences: the first in London, (July 2000); Amsterdam, (2001), Cardiff, (2002) and Dublin (2003). Our next conference will be in Klagenfurt, Austria, in July 2004. Conference sponsors have included ISACA, PWC and KPMG. EuSpRIG has recently been invited to give talks to the British Computer Society and auditors at the Bank of England.

## 2.  WHAT IS THE PROBLEM?

Spreadsheet risks are seldom mentioned in the media. When they are mentioned, however, it is usually with some sense of alarm as this report in the UK's Computer Weekly magazine suggests:

*'End users are putting their companies at risk by setting up spreadsheets without realising that this demands the discipline of traditional programming. Our [KPMG] findings are disturbing, but they are not really surprising, as 78% of models had no formal quality assurance to ensure they were built to specified requirements and were fit for purpose'* [Kavanagh J. 1997]

David Finch, Head of Internal Audit at Superdrug confirmed that some audit departments were aware of the problems associated with spreadsheets:

*"The use of spreadsheets in business is a little like Christmas for children. They are too excited to get on with the game to read or think about the 'rules' which are generally boring and not sexy"* [Chadwick D. 2000]

Public sector bodies have also shown an interest in the problem. Auditors from HM Customs & Excise have been concerned about spreadsheet integrity for several years. Their particular concern is that of spreadsheets used by businesses for making VAT declarations. It is common for VAT auditors to sample such spreadsheets and audit them thoroughly. Ray Butler of HM Customs & Excise has audited spreadsheets for many years and has written widely on the frequency of errors detected in VAT models. Butler has warned:

*"The presence of a spreadsheet application in an accounting system can subvert all the controls in all other parts of that system"* [Butler R. 2000]

At the 2002 EuSpRIG conference, Butler pointed out the use of a spreadsheet in the Allied Irish bank debacle two years ago [Butler R, 2002]. The fraud involved (amongst other things) falsification of spreadsheets used for monitoring the alleged perpetrator's work. These spreadsheets contained exchange rates ostensibly downloaded from Reuters on-line feed. The fraud resulted in the payment and potential payment of false staff bonuses of US$549,000 . Funds at risk totaled US$691.2 million.

*"Sarbanes-Oxley implies managers can't ignore un-controlled spreadsheets"* [Pettifor, B. Eusprig conference 2003]

At the 2003 EuSpRIG conference, reference was made to the failure to control a spreadsheet in the TransAlta energy company in Calgary, Canada. They lost $24M in June 2003 through a "cut-and-paste" error that mismatched prices.

Media reports such as the above attracted the attention of the early EuSpRIG researchers. They have been motivated to investigate further and collect evidence of risks, the frequency of risks and the possible solutions that may exist or may need developing.

## 3. SPREADSHEET DEVELOPMENT METHODS

One of the main problems of spreadsheet development is that no specific development methodology is available, unlike database construction which boasts such tools as ER diagrams and normalisation. Coupled with this is the fact that few organisations seem to encourage best practice, procedures and standards. Barry Pettifor, Director of Spreadsheet Assurance Services at PricewaterhouseCoopers, recently summed this up neatly:

*"We nearly always find that the modellers have no formal training in good modelling techniques, and that their organisations do not even have the most rudimentary internal modelling standards"* [Chadwick D. 2000]

EuSpRIG has made a point of encouraging work involving spreadsheet development methodologies that could reduce the frequency and impact of known risks. Several different approaches have been encountered. There have been appeals to a strict Software Engineering Approach, simple Life-Cycles to help training, Visual Aids to display formulae and constructs in less algebraic ways together with Programming Languages which can be compiled to form working spreadsheet models. Each of these approaches has been discussed in papers submitted to EuSpRIG conferences and they all have their strengths and weaknesses.

Thomas Grossman, of the University of Calgary, observes:

*"Spreadsheets are a powerful modeling language, mainly used by amateur programmers on a diversity of applications which are typically deployed throughout a wide range of different business functions"* [Grossman T. 2002].

Grossman argued that the basic tenets of software engineering best practice could be applied to the building of spreadsheets bearing in mind his own eight principles which state:

Principle 1: Best practices can have large impact
Principle 2: Lifecycle planning is important
Principle 3: A priori requirements specification is beneficial
Principle 4: Predicting future use is important
Principle 5: Design matters
Principle 6: Best practices are situation-dependent
Principle 7: Programming is a social, not an individual activity
Principle 8: Deployment of best practices is difficult and consumes resources

Jocelyn Paine at Bristol University, UK, has taken the software engineering analogy one step further by devising a spreadsheet modelling language called ModelMaster [Paine J. 2001]. ModelMaster has constructs similar to any programming language. Modelmaster uses named variables as column and row heading attributes as shown in this sample from one of his programs.

```
Attributes <
      new_quantity,  old_quantity,  new_real_income,  old_real_income,
   demand_change
   real_income_change, income_elasticity, good_type
➢Where
   demand_change = new_quantity / old_quantity – 1
   and real_income_change = new_real_income / old_real_income – 1
   and income_elasticity = demand_change / real_income_change
   and good_type =
      if (income_elasticity > 0,
         "normal",
         "inferior"
      )
```

When the ModelMaster program is compiled and executed it produces a typical spreadsheet matrix of rows and columns. In many ways, this fully addresses the software engineering analogy by starting with initial code and gains many merits for this. However, starting with actual code may lose some of the usefulness of the highly visual GUI approach of most spreadsheet software.

EuSpRIG will continue to support such approaches based upon the disciplines of software engineering as a promising line of action to pro-actively prevent problems occurring as models are developed.

## 4. SPREADSHEET AUDIT SOFTWARE

When audit tools are investigated it is soon noticeable how few organisations even consider auditing their spreadsheets. In one particular investigation in 1999, a City of London insurance company was found to have 18,000 spreadsheets on its office server. When staff were asked to archive (to a different server) files they no longer used, the number dropped to 17,300. This presumably indicated that staff still considered all these models to be of possible future use. The pattern of usage was discovered to be that a new model – a novel insurance risk for example – would be constructed from parts of past models by simply cutting and pasting. The frightening thing about this process was that, at no stage, were any of these models ever checked or audited for correctness.

Some of the risks of spreadsheets can be reduced by the use of audit tools providing there is a recognised organization-wide spreadsheet auditing function which often there is not.

However, there are several audit tools available that auditors may use: the built-in Microsoft Excel audit tool, Spreadsheet Professional, Excel Auditor, Spreadsheet Detective, (OAK) Operis Analysis Kit, SPACE (Spreadsheet Audit for Customs & Excise) and the Klagenfurt Tool Kit. Comparisons between some of these products have been made in a comprehensive paper delivered at EuSpRIG 2001 by David Nixon at Salford University [Nixon D. 2001].

The Microsoft Excel application has an audit tool built into it which, although rudimentary, can be useful in some circumstances. However, there is much anecdotal evidence that many spreadsheet builders have little knowledge that the audit tool even exists and even when they know of it, they rarely use it.

Customs & Excise have their own product based upon years of research into the types and causes of miscalculations found in spreadsheets used by UK businesses for calculating their VAT returns. This tool, called SpACE (Spreadsheet Audit from Customs & Excise) is a software add-in for Excel that tests *any* file Excel can open and works on a copy file so no alteration to the original can be ascribed to the test program (a legal pre-requisite). SpACE is part of a complete risk assessment and testing methodology and has been selling well in the UK, mainly to public-sector bodies.

## 6. EUSPRIG INITIATIVES

Since 1996 ISACF (Information Systems Audit & Control Foundation) and the IT Governance Institute have published CobiT (Control Objectives for Information & Related Technology) which brings mainstream IT control issues into the corporate governance arena. CobiT is a toolset to help business managers manage risks associated with implementing technologies, and demonstrate to regulators, shareholders, and other stakeholders how and how well they have done this. EuSpRIG has demonstrated that spreadsheet risks can be mapped on to the CobiT framework and so brought to managers' attention in a familiar format [Butler R. 2001]. EuSpRIG will be developing and publicising further endeavours in this area.

EuSpRIG has a keen interest in developing educational and training initiatives wherever spreadsheets are taught. David Banks and Ann Monday stated at the EuSpRIG 2002 conference:

*"As educators our concern is to try to develop the student skills in both the development of spreadsheets and in taking a critical view of their potential defects"*. [Banks D., Monday A. 2002].

To this end, EuSpRIG has become involved with organisations producing training materials for the ECDL (European Computer Driving License) and has been asked to comment upon forthcoming changes. The ECDL is a basic IT qualification for adults that has Europe wide recognition. It is hoped that this work will be the start of many training and educational inputs in the future. To assist with this, examples of common errors are now available on a Spreadsheet Model Database. This holds examples of real-world spreadsheets (suitably treated to protect confidentiality) containing errors, fraudulent statements etc., all of which may be used for research and teaching purposes for students, IT professionals and auditors.